\documentstyle[prd,aps,epsfig]{revtex}
\begin{document}
\author{Yuriy Mishchenko and Chueng-Ryong Ji}
\address{Department of Physics, North Carolina State University,
Raleigh, North Carolina 27695-8202}
\title{Molar mass estimate of dark matter from the dark mass distribution 
measurements}
\date{\today}
\maketitle
\tightenlines
\begin{abstract}
We study the distribution of dark matter versus visible matter using 
a set of data obtained from strong gravitational lensing in the galaxy 
cluster CL0024+1654 and another set of data inferred from the universal 
rotation curves in spiral galaxies. The important feature of these two 
dramatically different observations is that the mass density profile of 
both visible and dark components can be estimated. From these 
measurements we deduce the mass of the dark matter particle and
our estimate of the mass for the dark matter particle is 
$\mu_d \approx (200 \sim 800)$MeV. 
We contrast our estimates from CL0024+1654 data and  the universal 
rotation curves of the spiral galaxies and discuss their consistency.
\end{abstract}

\section{Introduction}
The problem of dark matter has been one of the 
biggest puzzles in astrophysics for more than a decade. 
First revealed in galaxy halos through the so-called rotation curves, it 
was later found in even larger scales such as in the clusters of galaxies. 
Nowadays, the dark matter is an important and intrinsic part of the 
cosmological picture of the universe. Even though nobody doubts the 
existence of this hidden substance these days, little unless nothing is 
known about its microscopic properties. Many candidate particles have been 
suggested for the composition of dark matter from various possible standard 
model extensions but their existence has not yet been confirmed 
experimentally.

In this paper, we are going to undertake a rather different approach and 
attempt to estimate the dark matter parameters from the available 
experimental data based on the simple but general physical principles. 
In the next section (Section II), we begin with the mass profile of 
CL0024 galaxy cluster measured by {\it Tyson et al.} in 1998 
\cite{tyson} and show a way for these data to reveal an amazing 
correlation between dark and visible mass distributions. 
Using a simple thermodynamic point of view, we relate the parameters of this 
correlation with the molar mass ratio of visible and dark matter.
In section III, we extend our thermodynamic approach to 
the case of spiral galaxies and analyze the Universal Rotation 
Curves (URC) as our input \cite{persic} to see if we can strengthen our 
conclusions. Remarkably, the estimate for the molar mass ratio in this 
dramatically different system comes out consistent with the result of 
section II. In section IV, 
we present possible interpretations of our findings and deduce that they 
indicate the mass of dark matter particle between 200MeV and 800MeV. 
Discussion and conclusion follow in section V. In Appendices A and B, the 
details of our URC analysis and the scaling properties of CL0024 mass 
profile are presented, respectively. 

\section{Analysis of the mass profile of galaxy cluster CL0024}
\label{secII}
The idea that the mass distribution in galaxies or galaxy clusters can be 
measured from the data on gravitational lensing is not new 
\cite{straumann}. In the last decade many systems exhibiting
gravitational lensing have been studied 
\cite{tyson,wu,kneib,abdelsalam,bezecourt,white}. 
Among them, the study by Tyson {\it et al.} in 
1998 \cite{tyson} is distinguished in that the detailed mass map was 
presented both for total and visible mass components. 
Assuming that the dark matter follows the known thermodynamical laws,
we now note that the comparison of these two maps may give a valuable 
insight in properties of dark matter particles from the experimental
perspective.

A brief survey of the measurement is in place.
In 1998 the study of Hubble telescope images of galaxy cluster CL0024+1654 
has been carried out by Tyson {\it et al.} and the mass
density profile of the cluster has been derived from strong gravitational 
lensing \cite{tyson}.
It was found that the vast majority of the mass is not associated
with the galaxies and forms a smooth elliptical distribution,
slightly shallower than isothermal sphere, with a soft core 
of $r_{core}=35\pm3 h^{-1} kpc$, where $h$ is the normalized Hubble
constant. No evidence of in-falling massive clumps has been found 
for the dark component.
The projected dark matter density profile is well fit by a power law model
\begin{equation}
\Sigma(y)=\frac{K(1+\eta y^2)}{(1+y^2)^{2-\eta}},
\end{equation}
where $y=r/r_{core}$, $K=7900\pm100 h M_\odot pc^{-2}$, 
$r_{core}=35\pm3 h^{-1}kpc$ and $\eta=0.57\pm0.02$\cite{tyson,shapiro}. 
The primary conclusion from the observed mass distribution was 
that the predicted profile in Ref.\cite{navaro} was 
inconsistent with the observed result.

In this section, we attempt to study the mass density profile in CL0024 
from a thermodynamic point of view. 
In particular, we assume that the dark matter complies with the known 
thermodynamic laws and that it is in the classical region so that 
Boltzmann statistics can be applied. 
If the rotation can be neglected, 
the general thermodynamics principles
imply 
the following distribution for 
visible and dark mass components in a gas cloud 
\cite{hatsopoulos}
\begin{equation}
\label{eqn002z}
\begin{array}{c}
\rho_i (r) = a_{i} e^{-\beta_i(r)\mu_i \Phi(r)} , \cr
\end{array}
\end{equation}
where $\mu_i$ is the molar mass for the corresponding component, 
$\beta_i=1/kT_i$ and 
$\Phi(r)$ is the gravitational potential
at position $r$.
Even though the details of the radial distribution of the gravitating gas
may be difficult to model and understand 
\cite{votyakov,vega1}
(see {\it e.g.} Appendix 
\ref{appB} 
for some discussion on the radial mass distribution in CL0024), 
Eq.(\ref{eqn002z})
leads to a simple relation between mass densities of 
dark and visible components, {\it i.e.}
\begin{equation}
\ln \rho_i(r) - d_i = -\mu_i \beta_i(r) \Phi(r),
\end{equation}
where $d_i = \ln a_{i}$.
One may expect therefore to see
a linear correlation between $\rho_v(r)$ and $\rho_d(r)$ by
plotting the data on Log-Log scale. The slope $\kappa$ is then given by 
the ratio of the Boltzmann factors for visible and dark matter, {\it i.e.} 
$(\kappa\approx \beta_v\mu_v)/(\beta_d \mu_d)$.
Although the local fluctuations in the mass density of one component
might locally shift $\ln( \rho_i(r) )$, outside of the fluctuation 
region we would still expect to see a line with a certain slope. 
This motivates us to study these distributions in CL20024 from the 
assumption that elementary thermodynamics laws can be applied to the 
analysis of the mass density profile including the dark matter halos. 
Note that the "linear regime" in Log-Log plot of visible vs. dark matter
distribution can be easily spoiled by many factors, 
e.g. non-equilibrium processes, local fluctuations or effects of rotation.
Thus, the validity of this thermodynamic assumption must be carefully
evaluated from the consistency of the results obtained from the
analysis of different systems, as we discuss in the present and 
next sections (Section II and III).

\begin{figure}
\centering
\epsfig{file=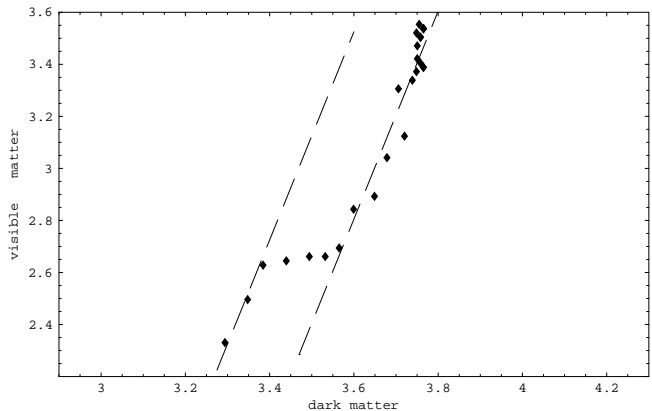,width=250pt}
\caption{Data set extracted from mass profiles of 
CL0024\protect\cite{tyson}.}
\label{dataset}
\end{figure}
In  Fig.\ref{dataset}, we present the sample of points taken from the graph 
for the radial projected mass distribution of dark and visible matter
by Tyson {\it et al.} \cite{tyson} in the Log-Log scale.
In the plot we indeed observe a linear 
correlation. The short horizontal segment on the graph, 
once related to the radial mass distributions from Tyson {\it et al.}, 
could indicate presence of two major subsystems in the 
visible component
of the cluster: 
inner with smaller spread, and outer with larger spread. 
We can attribute the inner subsystem to the region of 
gravitational influence of the central formation in the cluster, while the 
outer subsystem corresponds to the concentrations of matter formed 
outside of this region.
Transition between the two subsystems occurs in relatively narrow region 
at distance of about 100 $h^{-1}$ kpc, as can be observed from 
the original plots by Tyson {\it et al.}, and manifests itself as the 
above mentioned flat segment.
While this effect is rather interesting and deserves further attention in 
its own, we will leave it aside for the sake of main point in the present
discussion.

Outside of the transition region, remarkably, the datum reveals linear 
correlations with the same slopes.
To extract this slope we may further deal with the dataset by 
either cutting off all points beyond the bridge or by cutting out only 
the transition region and
keeping the other points in the set:
\begin{figure}[htb]
\centering
\begin{minipage}[c]{0.45\hsize}
\epsfig{file=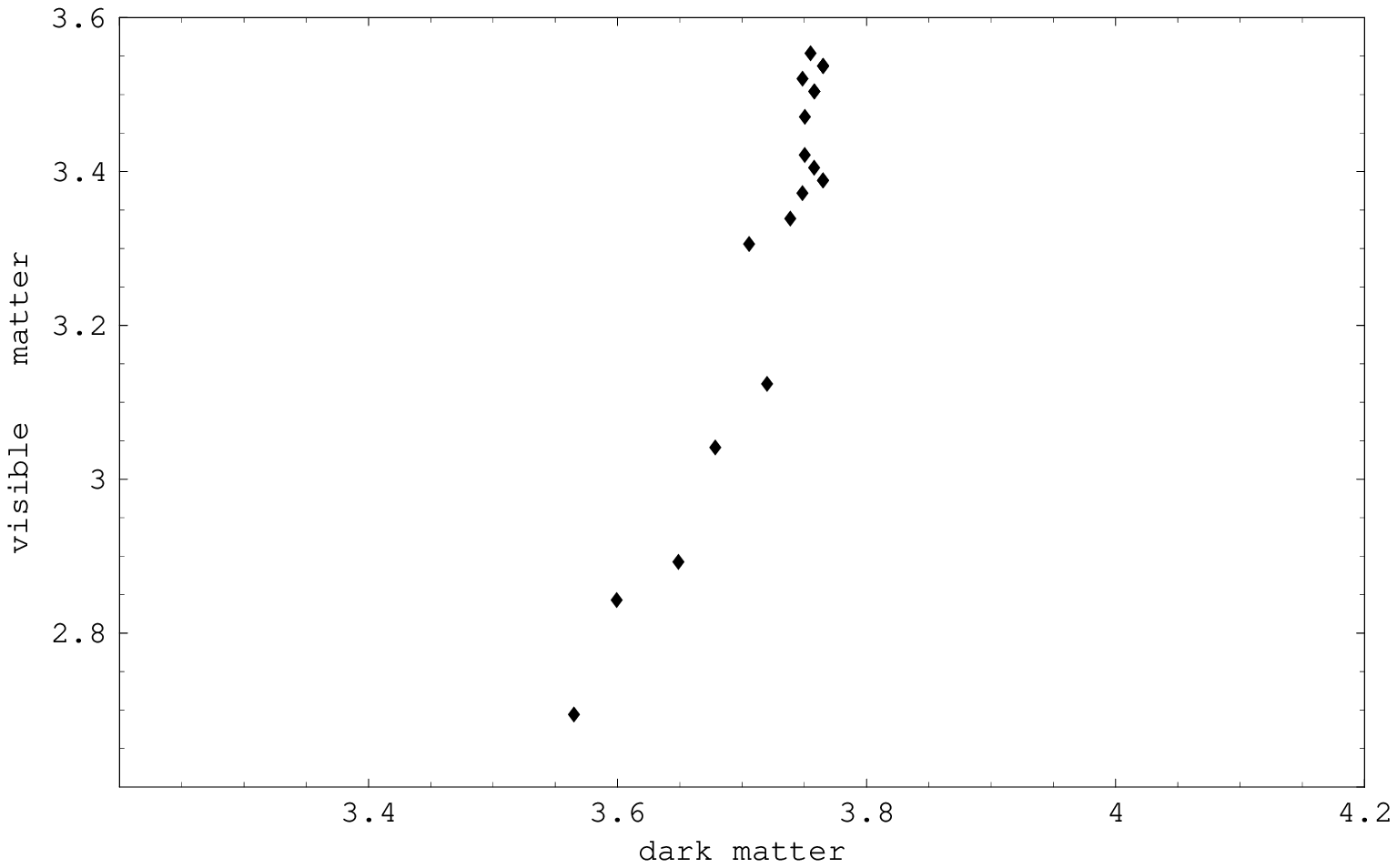,width=\hsize}
\end{minipage}
\hspace*{0.5cm}
\begin{minipage}[c]{0.45\hsize}
\epsfig{file=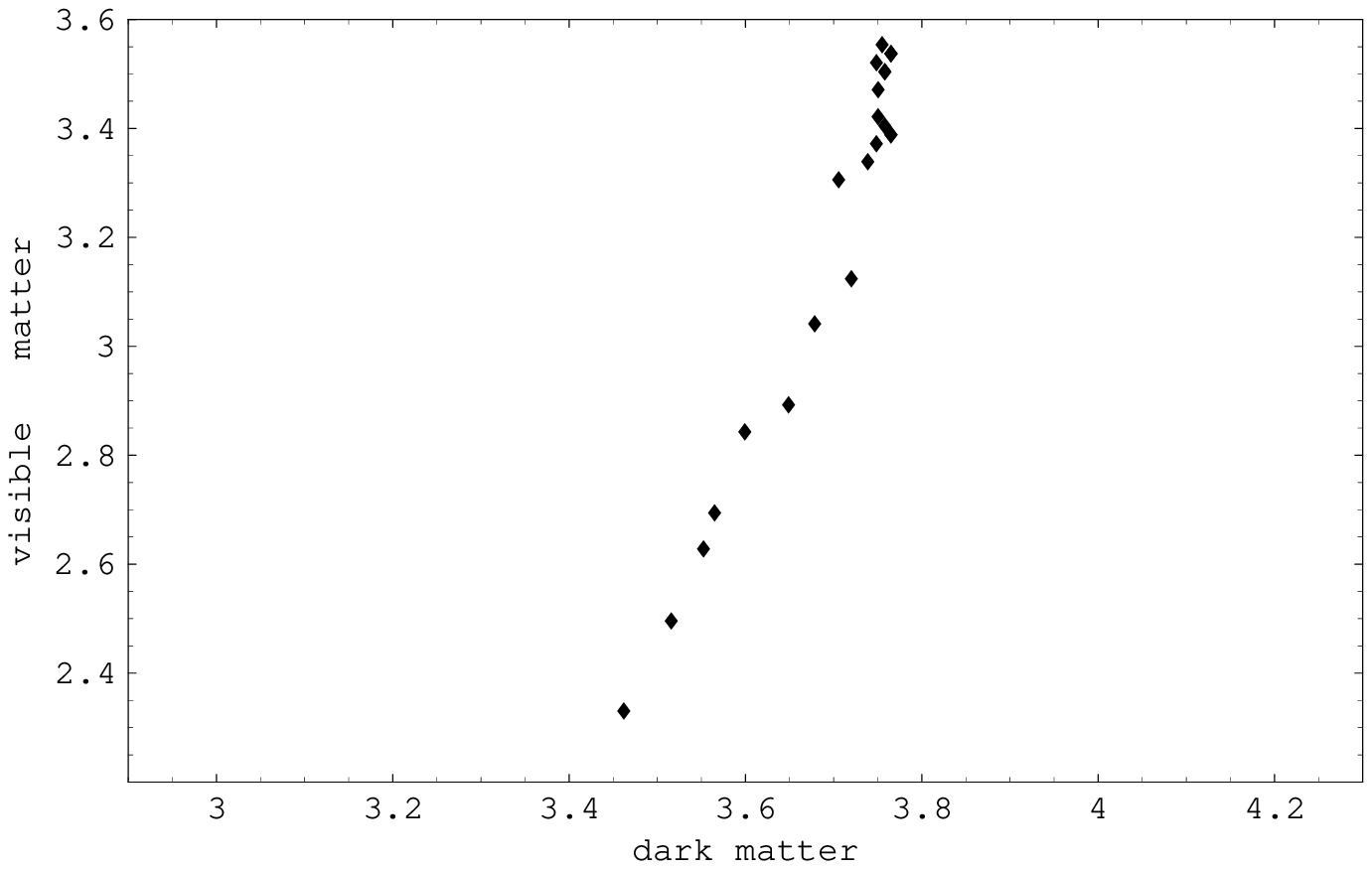,width=\hsize}
\end{minipage}
\caption{Truncated and corrected datasets from the graph by Tyson {\it 
et al.}}
\label{linears}
\end{figure}
In both cases, indeed, the estimate for the slope of the correlation is 
similar. For the truncated set one obtains $\kappa\approx (3.55-4.9)$ 
while for the corrected set one obtains $\kappa\approx (3.61-4.35)$. Guided by the above thermodynamical idea, we can conjecture from these plots that the ratio of the Boltzmann 
factors for dark mass distribution and visible mass distribution is 
consistent with a constant given by $\kappa \approx (3.6-4.4)$.
\begin{figure}
\centering
\epsfig{file=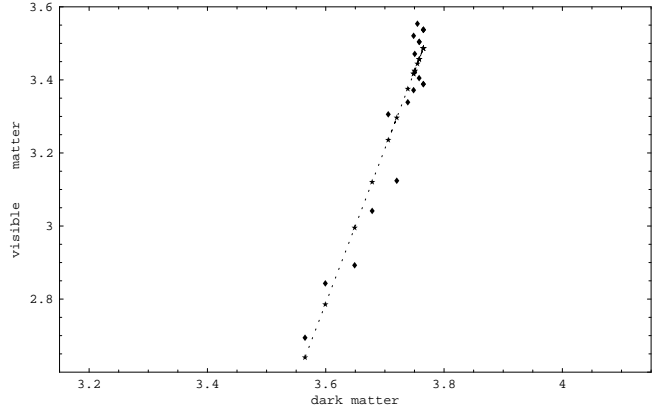,width=250pt}
\caption{Linear fit to the data presented in Fig.\protect\ref{linears}.}
\label{fit}
\end{figure}

\section{Analysis of the Universal Rotation Curves for spiral 
galaxies}\label{secIII}

\begin{figure}[htb]
\centering
\begin{minipage}[c]{0.45\hsize}
\epsfig{file=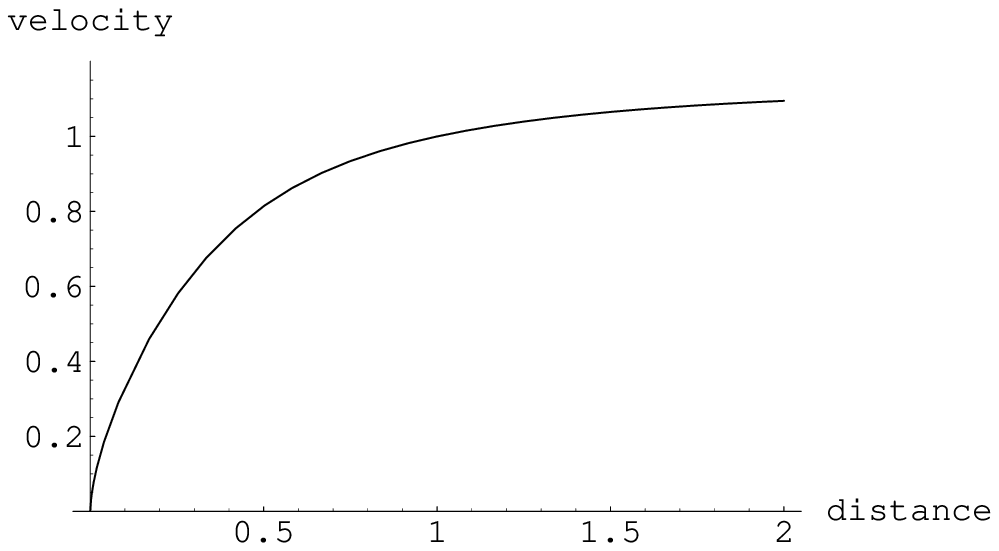,width=\hsize}
\end{minipage}
\hspace*{0.5cm}
\begin{minipage}[c]{0.45\hsize}
\epsfig{file=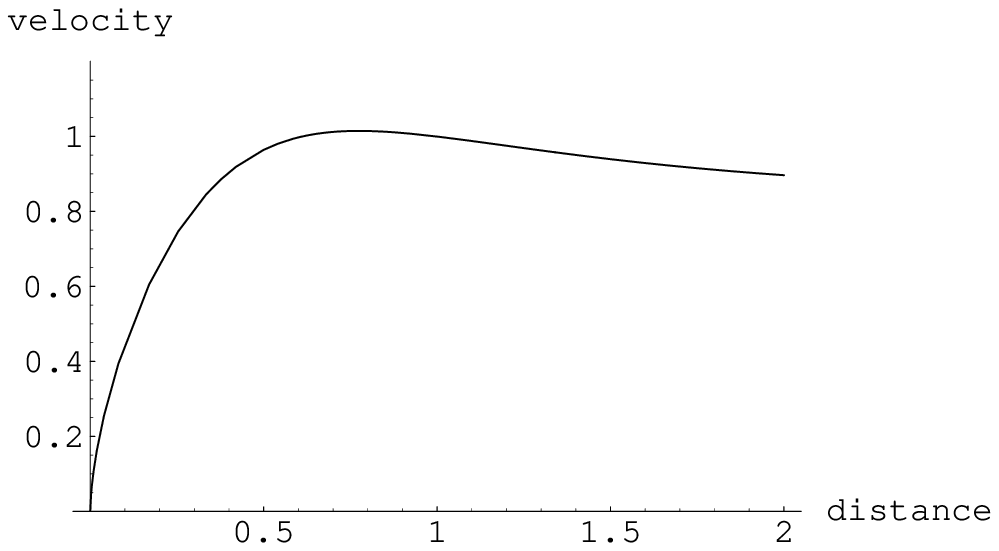,width=\hsize}
\end{minipage}
\caption{URC for low luminosity galaxy (M=-18.5, left) and high luminosity
galaxy (M=-23.2, right). Velocity is normalized to $v_{opt}$, distance is 
normalized to $r_{opt}$. See the text for the notations $v_{opt}$ and 
$r_{opt}$.}
\label{urcvel}
\end{figure}

It is well known that the rotation curves (RC) of spiral galaxies do not
show any Keplerian fall-off beyond their optical radius (see 
Fig.\ref{urcvel}). This is one of the most
remarkable manifestations of large dark mass component in galaxies. 
Thus, RCs are indeed the measurements of dark mass distribution in
galaxies and it would be interesting to see if the approach 
presented in Section \ref{secII} yields 
a meaningful result in this case.
We shall also note that, although the measurement of the mass distribution 
in CL0024 cluster is a remarkable result, it is quite unique in the 
sense that there are just few other studies that yield detailed 
distribution of dark vs. visible matter in galaxy clusters. 
On the other hand, RCs have been studied for a long time in many galaxies 
(over 1000 by now) and thus can provide a good consistency check for our 
method. Moreover, the spiral galaxies and the galaxy clusters are 
tremendously different systems in their scales and dynamics. 
The consistency of our approach in these two dramatically different 
regimes would provide a good evidence of its validity.

Application of the above ideas to the mass distribution in spiral 
galaxies, however, faces many experimental as well as theoretical 
complications. It is experimentally difficult and controversial to 
measure RC and map out the visible mass contribution to it 
\cite{combes,albada}.
Although a large number of such separations have been carried out 
\cite{persic}, 
the quality of these results is uncertain. 
Furthermore, the measured RCs rarely extend to more than 1.5-2 optical 
radius. Thus, in most cases, only a very limited part of dark matter 
distribution becomes visible that doesn't include the fall-off tail which is the most sensitive part for the properties of dark matter. 
Also, theoretically, the spiral galaxy is a complex conglomerate of dark 
matter, interstellar gas, stars and radiation in the presence of essential 
gravitational pull and rotation. Thus, it is unclear if such a system 
should satisfy the thermodynamic Boltzmann distribution at all.
Even if it does, how the effect of rotation should be included into 
the analysis is not absolutely clear. Finally, the global temperature 
variations along the stellar disk may pose a serious problem for the 
analysis.

In our study, we therefore should check first the consistency of the 
Boltzmann distribution in spiral galaxies. To begin our study, we 
accept ``the exponential rotating disk plus spherical dark halo" as a 
conventional mass model for the spiral galaxy and the Universal Rotation 
Curves (URC) \cite{salucci} as the experimental input for RCs (see 
Appendix \ref{appA}).

\begin{figure}[htb]
\centering
\begin{minipage}[c]{0.45\hsize}
\epsfig{file=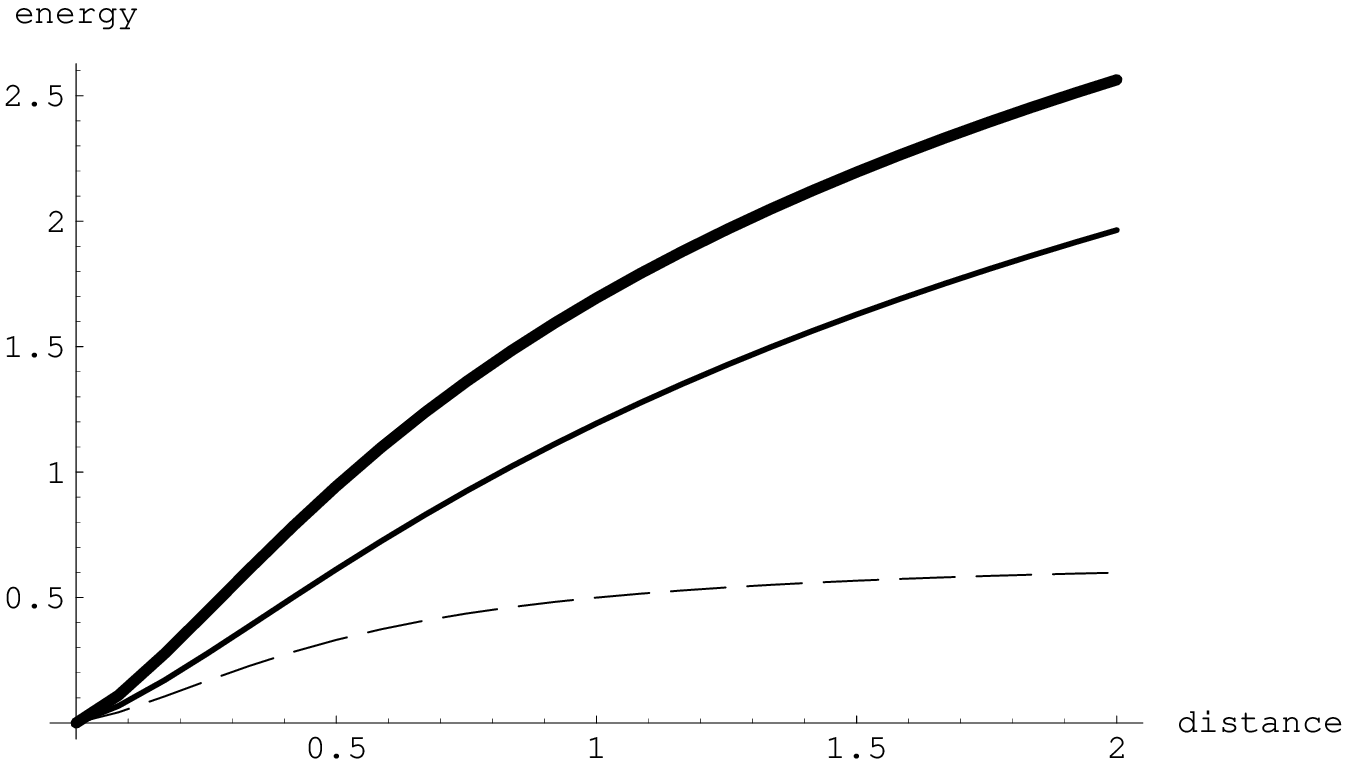,width=\hsize}
\end{minipage}
\hspace*{0.5cm}
\begin{minipage}[c]{0.45\hsize}
\epsfig{file=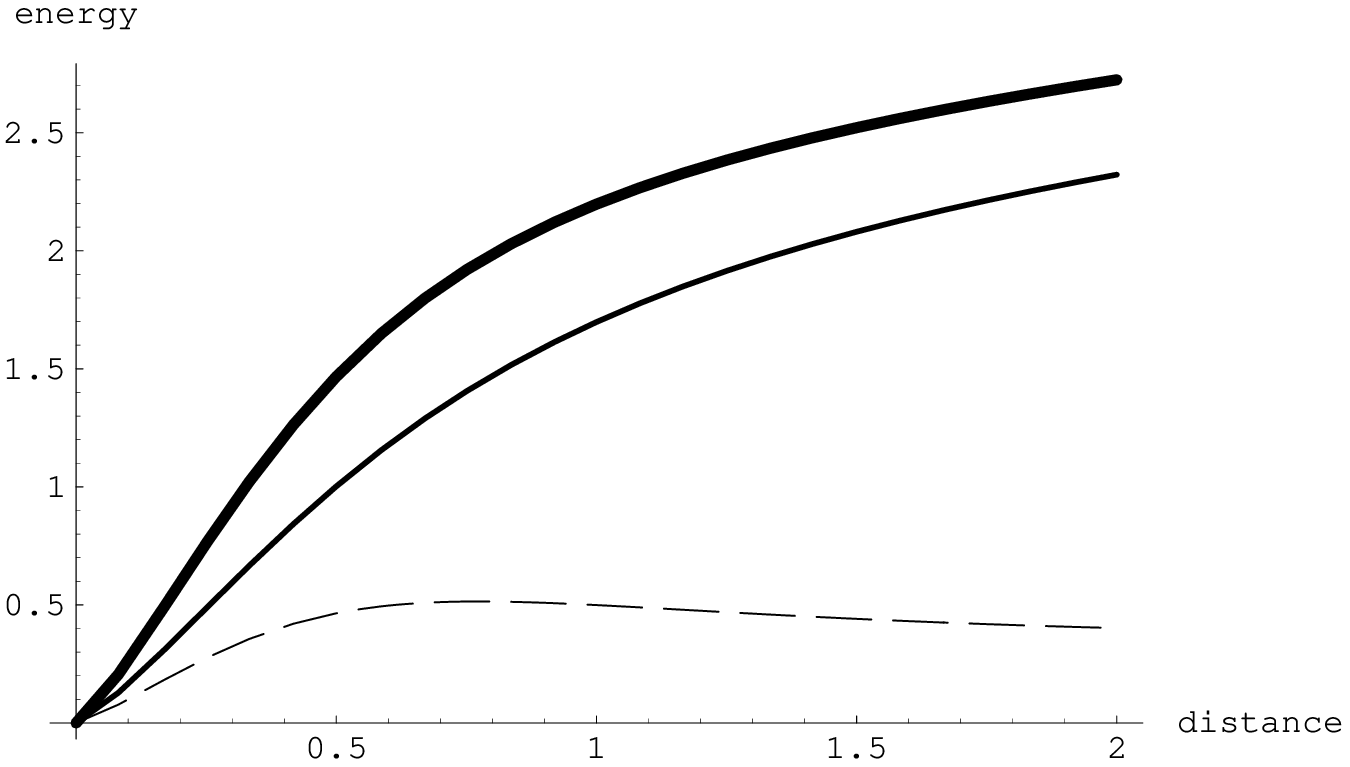,width=\hsize}
\end{minipage}
\caption{Total (thick line), gravitational potential (solid line) and 
kinetic energy (dashed line) in a spiral galaxy mapped out from URC. 
The left and right plots are for M=-18.5 and M=-23.2 luminosities, 
respectively.}
\label{urcpot}
\end{figure}

Note that URC maps out the gravitational potential throughout the galaxy 
disk via 
a simple relationship
\begin{equation}
v_{URC}^2(r)/r= \hat{r}\cdot\vec{\nabla} \Phi(r) = \frac{d\Phi(r)}{dr},
\end{equation}
so that for the Boltzmann distribution of mass we can write
\begin{equation}
\begin{array}{c}
\rho_v (r) = a_v e^{-\beta_v \mu_v E(r)}, \\
E(r)=\Phi(r) + v_{URC}^2(r)/2, \Phi(r)=\int\limits_0^r dr' v_{URC}^2(r')/r',
\end{array}
\end{equation}
where $\beta_v = 1/kT_v$, $\mu_v$ is the molar mass of the matter and 
$E(r)$ is the specific energy (by mass) at distance $r$ including 
the contribution from 
the rotation which we assume in the form $v_{URC}^2(r)/2$. 
Note that, owing to the spherical form of the dark halo, we may assume that 
the rotation of the dark halo is negligible, {\it i.e.}
\begin{equation}\label{eqn01e}
\rho _d (r)=a_d e^{-\beta_d \mu_d \Phi(r)}. 
\end{equation}
In Fig.\ref{urcpot}, we present the radial distribution of the gravitational 
potential and rotation energy, normalized to $v_{opt}^2 = 
v_{URC}^2(r_{opt})$, vs. the distance, normalized to the optical 
radius denoted by $r_{opt}$. From this figure, one can note that the 
relative effect of rotation decreases with distance and reduces to almost 
constant contribution at $r\gtrsim 0.6 r_{opt}$. 
The effect of rotation on the Boltzmann distribution is therefore negligible at these distances.

\begin{figure}[htb]
\centering
\begin{minipage}[c]{0.45\hsize}
\epsfig{file=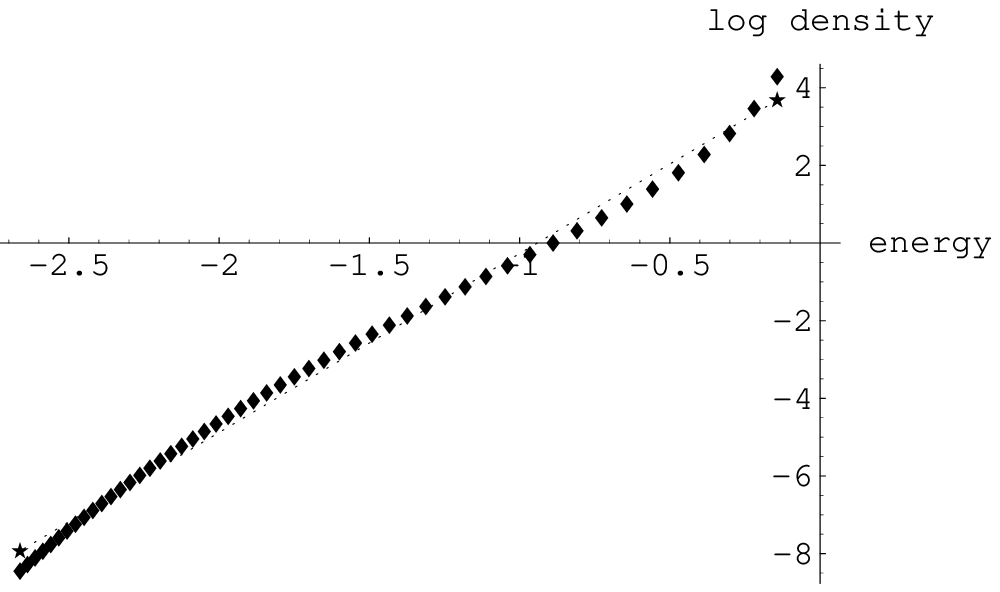,width=\hsize}
\end{minipage}
\hspace*{0.5cm}
\begin{minipage}[c]{0.45\hsize}
\epsfig{file=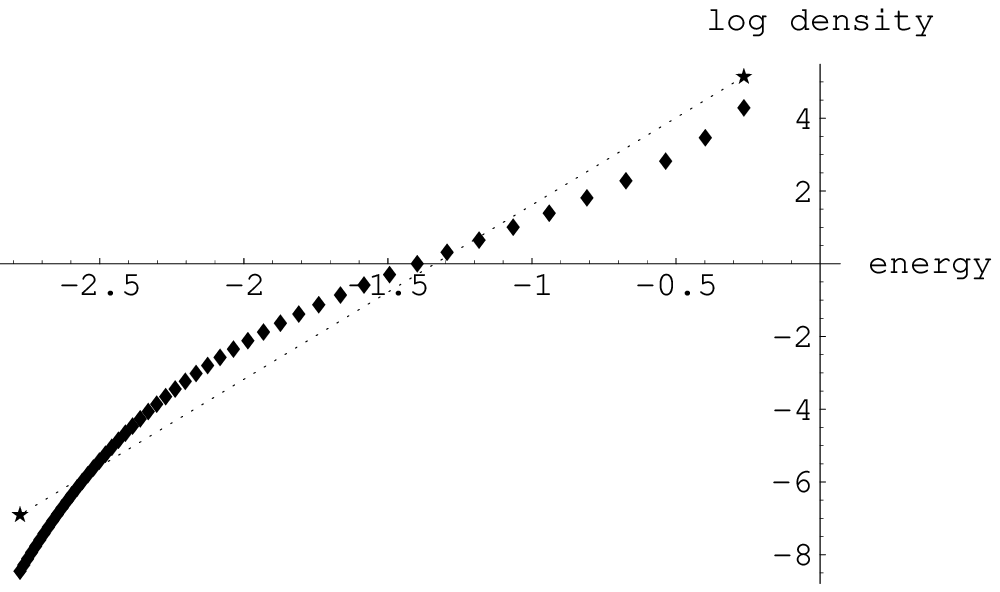,width=\hsize}
\end{minipage}
\caption{Visible mass distribution vs. energy for M=-18.5(left) and 
M=-23.2(right).}
\label{urcve}
\end{figure}

\begin{figure}[htb]
\centering
\begin{minipage}[c]{0.45\hsize}
\epsfig{file=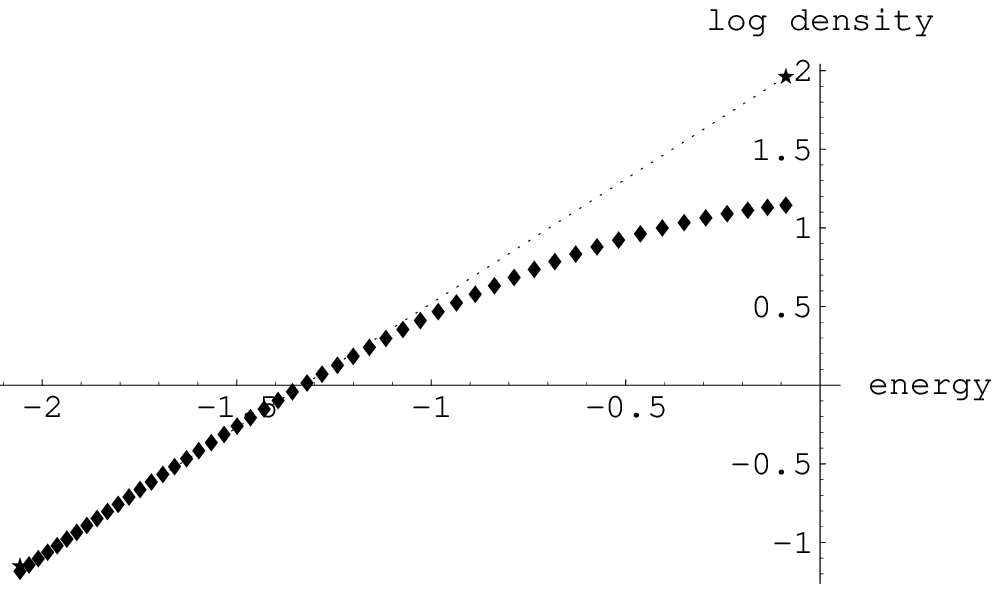,width=\hsize}
\end{minipage}
\hspace*{0.5cm}
\begin{minipage}[c]{0.45\hsize}
\epsfig{file=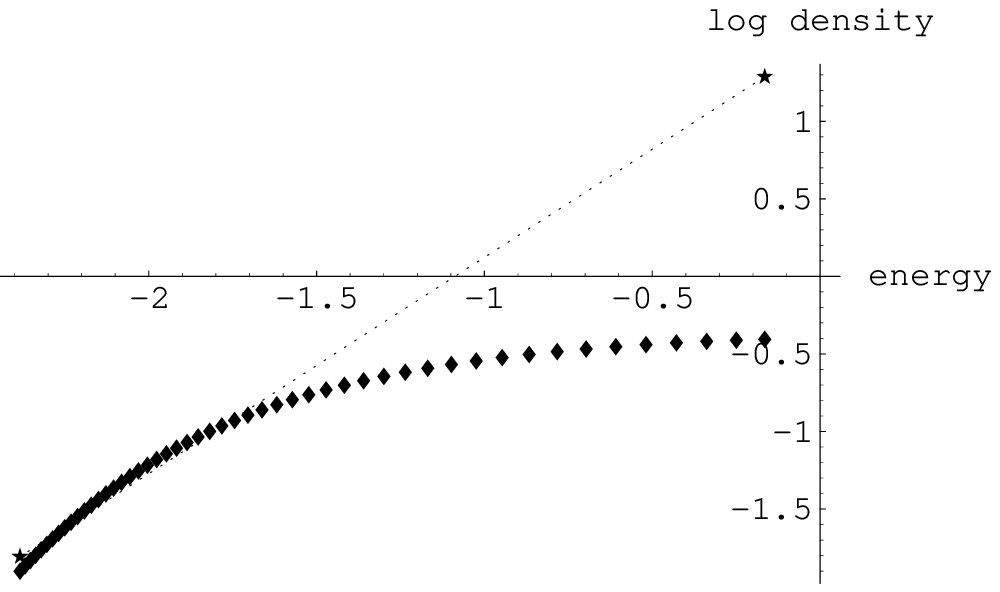,width=\hsize}
\end{minipage}
\caption{Dark mass distribution vs. energy for M=-18.5 (left) and 
M=-23.2(right).}
\label{urcde}
\end{figure}

\begin{figure}
\centering
\epsfig{file=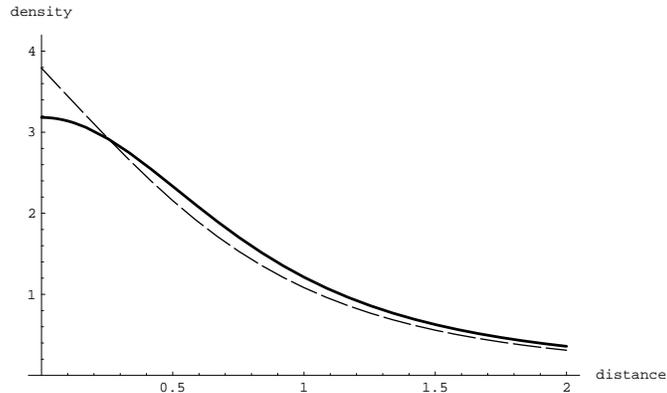,width=250pt}
\caption{Two sample density profiles that give perfect fits to URC.}
\label{urcden}
\end{figure}

To check the consistency of the Boltzmann distribution, we should compare 
consequently $\ln(\rho_v)$ with $E(r)=\int\limits_0^r dr' v_{URC}^2(r')/r' + 
v_{URC}^2(r)/2$ and $\ln(\rho_d)$ with $\Phi(r)=\int\limits_0^r dr' 
v_{URC}^2(r')/r'$, respectively.
Despite many possible complications mentioned above, 
for the distribution of visible mass, we do find a good linear 
correlation between $\ln(\rho_v)$ and $E(r)$ for the entire range of 
available URCs as shown in Fig.\ref{urcve}. The correlation gets worse 
for galaxies with higher luminosities 
which we may attribute to increasing
global variations of temperature along the stellar disk and increased 
effects from the galaxy bulge.
For the dark mass, as shown in Fig.\ref{urcde}, we observe much worse 
situation since the linear regime 
is reached only in the outer part of the stellar disk $r \gtrsim 0.6 
r_{opt}$.
While this might be used to argue against our assumptions, we must also 
note that the dark matter distribution is usually inferred from the 
separation of RC and thus is subject to a large uncertainty. 
This is quite different from the case of visible mass distribution which is 
modeled by the light/mass ratio from the photometric data.
Indeed, in Fig.\ref{urcden}, we 
compare the two quite different dark mass distributions, used in 
\cite{salucci} and \cite{persic}, {\it i.e.} solid and dashed 
lines, respectively. Both give perfect URC fits but they behave quite 
differently in small $r$ region.
We must therefore conclude that the URC is not sensitive to the details of 
small $r$ behavior of the dark mass distribution and the discrepancies in 
Fig.\ref{urcde} cannot be used to ultimately discard the validity of 
Boltzmann distribution in dark halos. Instead, we should check if 
an acceptable fit to URC can be made by assuming the dark mass 
distribution in the form given by Eq.(\ref{eqn01e}).

We performed such fits for a wide range of luminosities from M=-18.5 to 
M=-23.2 and found that a perfect fit can be constructed by varying $a_d$ 
and $\kappa_d=\beta_d \mu_d$ in every case that we considered.
\begin{figure}[htb]
\centering
\begin{minipage}[c]{0.45\hsize}
\epsfig{file=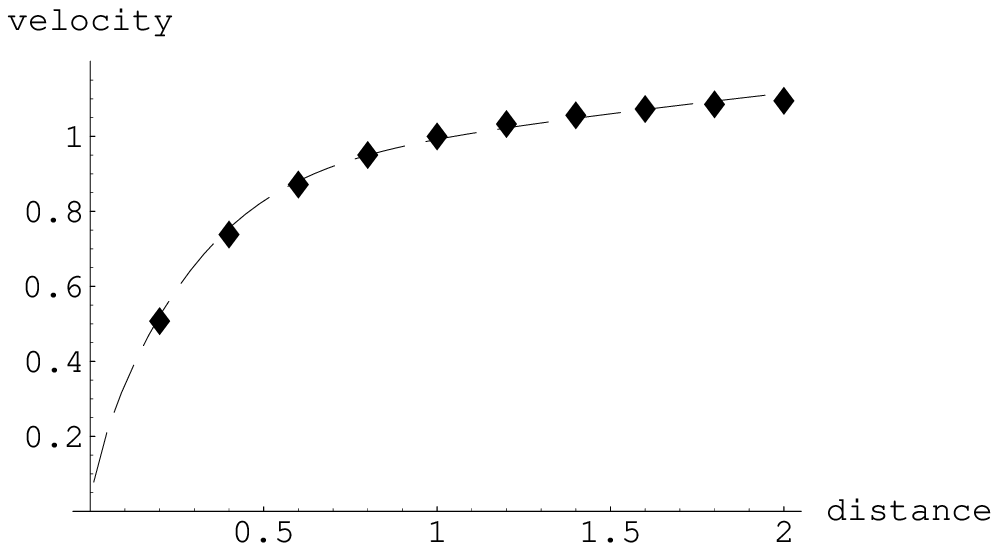,width=\hsize}
\end{minipage}
\hspace*{0.5cm}
\begin{minipage}[c]{0.45\hsize}
\epsfig{file=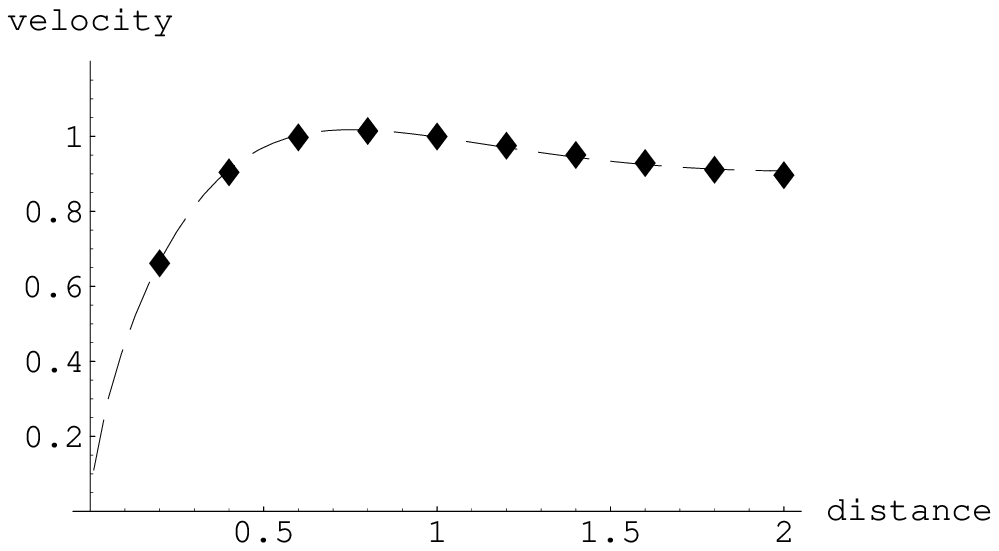,width=\hsize}
\end{minipage}
\caption{Boltzmann distribution fit to URC for M=-18.5 (left) and 
M=-23.2 (right).} 
\label{urcfit}
\end{figure}
We also found that, although the best fit $\kappa_d$ for each luminosity M varies with M ($\kappa_d\approx 0.8 - 1.2$), the URC is
not very sensitive to $\kappa_d$ and very good fits can be obtained 
in all cases even if $\kappa_d$ is kept constant and 
{\em only $a_d$ is varied}.
\begin{figure}[htb]
\centering
\begin{minipage}[c]{0.3\hsize}
\epsfig{file=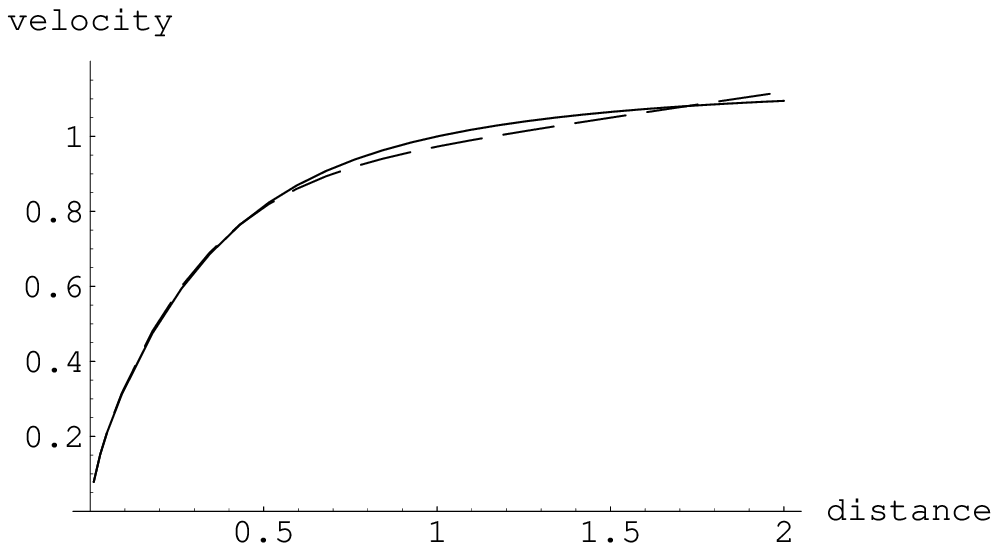,width=\hsize}
\end{minipage}
\hspace*{0.5cm}
\begin{minipage}[c]{0.3\hsize}
\epsfig{file=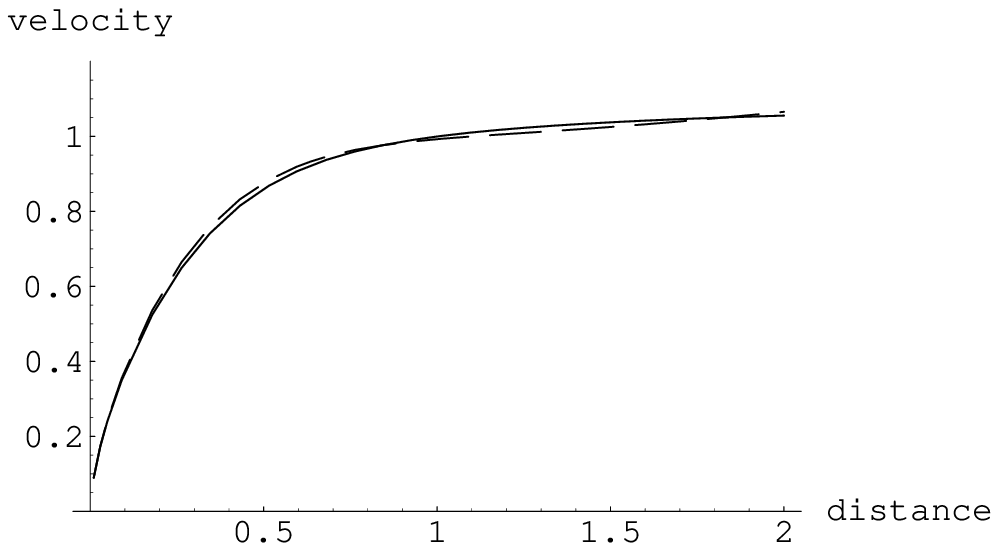,width=\hsize}
\end{minipage}
\hspace*{0.5cm}
\begin{minipage}[c]{0.3\hsize}
\epsfig{file=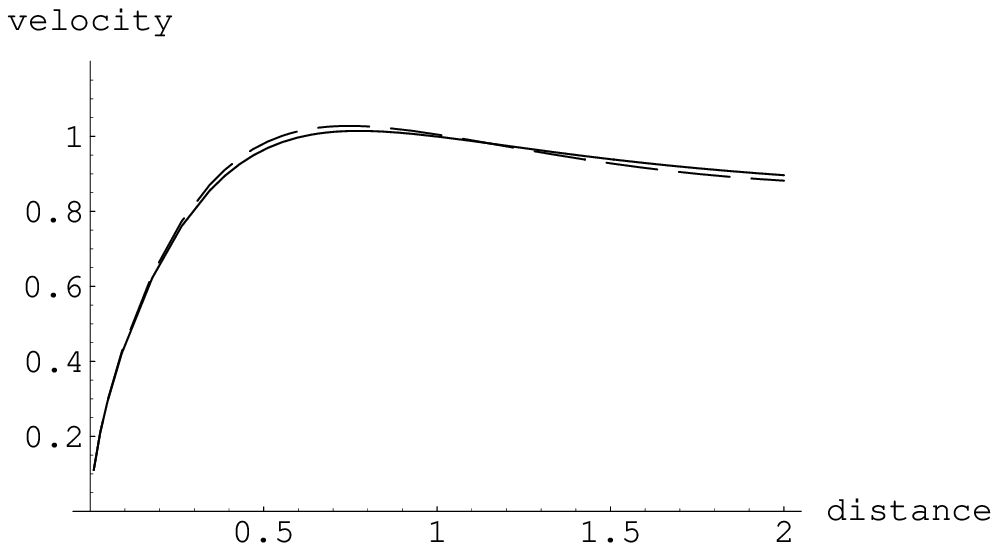,width=\hsize}
\end{minipage}
\caption{Boltzmann distribution fit (dashed line) with $\kappa_d\approx 
1.1$ kept constant for low (M=-18.5, left), intermediate (M=-20, center) and high (M=-23.2, right) luminosity galaxy.}
\label{urcfits}
\end{figure} 
From the above discussion, we can conclude therefore that the global 
profile of visible and dark mass in spiral galaxies is consistent with almost 
isothermal Boltzmann distribution.
Even though the data of visible and dark mass distribution plotted in
Figs.\ref{urcve} and \ref{urcde} only partially justify the use of
the linear regime, we have found that it is satisfactory in explaining
the URCs plotted in Figs.\ref{urcfit} and \ref{urcfits}.
The temperature of the distribution can be estimated as  
\begin{equation}\label{temp}
T\sim \frac{\mu_{H_2} v_{opt}^2}{\kappa_v k_B}\sim 
10^{5} - 10^{6} $K$,
\end{equation}  
which is consistent with the temperature of the interstellar gas.

Consequently, we found that the ratio $\kappa_v/\kappa_d\approx 5$ for a 
range of galaxy luminosities. Since we used a synthetic URC as our input 
data and a synthetic mass model for the visible mass distribution, it is 
impossible to properly estimate the error in $\kappa$
and we assume the variance of $\kappa$ with M in our URC-fits, 
e.g. $\kappa_d\approx 0.8-1.2$, as such an estimate.
In this case 
$\kappa_v \approx 4.7\pm0.1$ 
and $\kappa_d \approx 1.0\pm 0.2$, from which we obtain the ratio
\begin{equation}
\kappa = (3.7- 5.7).
\end{equation}

Our result here is consistent with analysis of the CL0024 galaxy cluster. 
This is quite remarkable since there is a  tremendous difference in the 
systems under consideration. A lot of complexities in dynamics of spiral 
galaxies might have affected our thermodynamic picture. Nevertheless, 
in both situations we may conclude that the ratio of 
the Boltzmann factors for visible and dark matter is a constant. 
Including both of our analysis presented in Sections 
\ref{secII} and \ref{secIII}, the constant can be estimated as $\kappa = 
3.6 \sim 5.1$.

\section{Interpretation of the result}
As was already mentioned in Section \ref{secII}, 
$\kappa$ can be related to the ratio of molar masses for the visible and 
dark components $\kappa = (\beta_v / \beta_d) (\mu_v / \mu_d)$.
In the previous sections(Sections II and III), we found that this ratio is 
constant in a variety of
very different conditions such as the galaxy clusters and 
the spiral galaxies of different luminosities.
Although it is still an open question how the two apparently different
systems, {\it i.e.} dark matter and visible matter, could share the same
temperature, it is essentially the local thermodynamic equilibrium
which we think is the most natural way to consistently describe the data
with such variety of different conditions.
Since the temperature ($\beta$) and the molar mass ($\mu$) are independent
variables, it is unlikely that $\mu_v/\mu_d$ varies in a precise 
correlation with $\beta_v/\beta_d$ in such a way that $\kappa$ is a 
constant both for the CL0024 galaxy cluster and the spiral galaxies
with various luminosities. This leads rather naturally that 
$\beta_v/\beta_d$ must be a constant within about 20\% from our analyses 
presented in Sections II and III. Moreover, the temperature is likely
to vary with the distance as shown in Appendix B but yet no significant 
change in $\kappa$ is noticeable in the mass profiles of CL0024. 
Also, the temperature given by Eq.(\ref{temp})
vary widely even though $\kappa_v$ and
$\kappa_d$ vary insignificantly as shown in Section III,
since it is proportional to $v_{opt}^2$, yet the ratio $\beta_v/\beta_d$
is consistent with constant.
Due to such correlation between $\beta_v$ and $\beta_d$ 
evident in our analysis,
we think that the local thermodynamic equilibrium between the dark and
visible matter, {\it i.e.} $\beta_d(r)\approx \beta_v(r)$, is the
most natural way to describe these data.
This might imply that the gravitational scattering 
in the dark matter 
may create non-vanishing
relaxation effect at the scales and times of the galactic halos.   
Feasible mechanism to reach the local thermal equilibrium along
this line may deserve further investigations.
Under the local thermodynamic equilibrium between
dark and visible matter
\begin{equation}\kappa \approx \mu_v/\mu_d. 
\end{equation}
Hence, our result gives us a possibility to estimate $\mu_d$ once $\mu_v$ 
is known. Since the ordinary matter primarily consists of $H$, 
$H_2$ and 
$He$ \cite{thronson}, 
we immediately conclude that the range for the mass of dark matter particle 
is from 150MeV to 1250MeV. 
The exact number, however, may depend 
significantly on the exact composition of visible matter. 

\begin{figure}
\centering
\epsfig{file=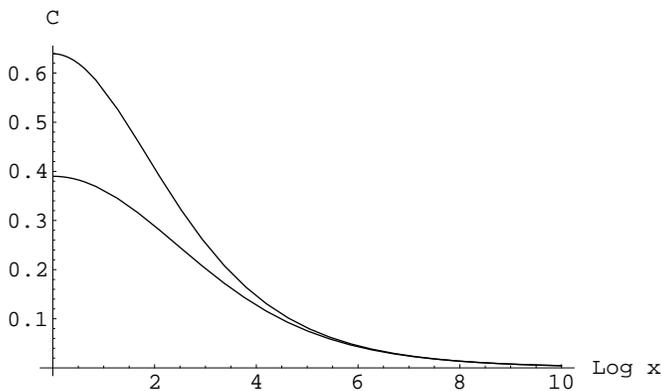,width=250pt}
\caption{Relative content of two-component gas in strong gravitational 
field.} 
\label{rel_vd}
\end{figure}

Indeed, if one takes into account that the visible matter is made of 
few components which are in fairly strong gravitational field,
one finds that the relative content of matter in such conditions 
changes dramatically with the distance ({\it e.g.} In 
Fig.\ref{rel_vd}, the upper and lower lines represent the densities of $H_2$ 
and $He$, respectively, as functions of the logarithmic distance; they 
are normalized to 1 at the origin) \cite{vega}. 
One may ask then if the linear correlation is at all to be expected 
between $\ln ( \rho_d )$ and 
$\ln ( \rho_v ) = \ln(\rho_{H_2}+\rho_{He})$. 
To answer this question we perform a simulation for such situation with 
different relative contents for visible mass.
\begin{figure}[htb]
\centering
\begin{minipage}[c]{0.3\hsize}
\epsfig{file=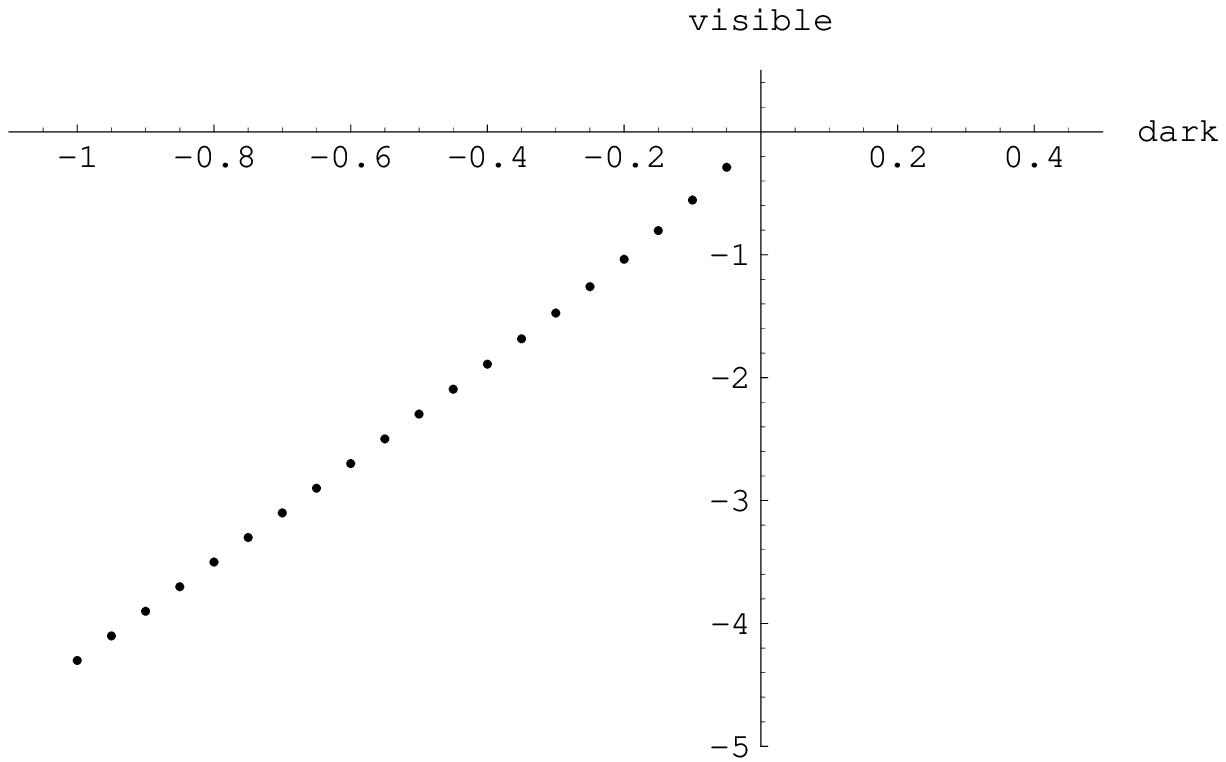,width=\hsize}
\end{minipage}
\hspace*{0.5cm}
\begin{minipage}[c]{0.3\hsize}
\epsfig{file=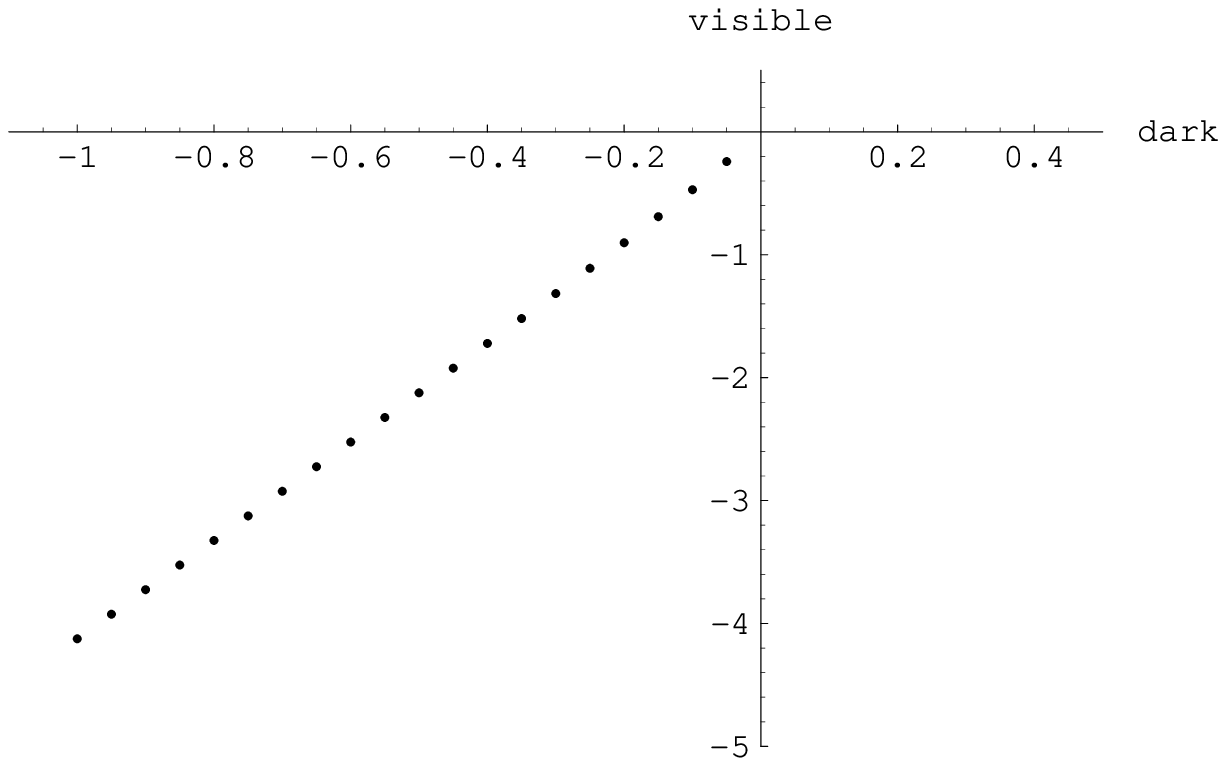,width=\hsize}
\end{minipage}
\hspace*{0.5cm}
\begin{minipage}[c]{0.3\hsize}
\epsfig{file=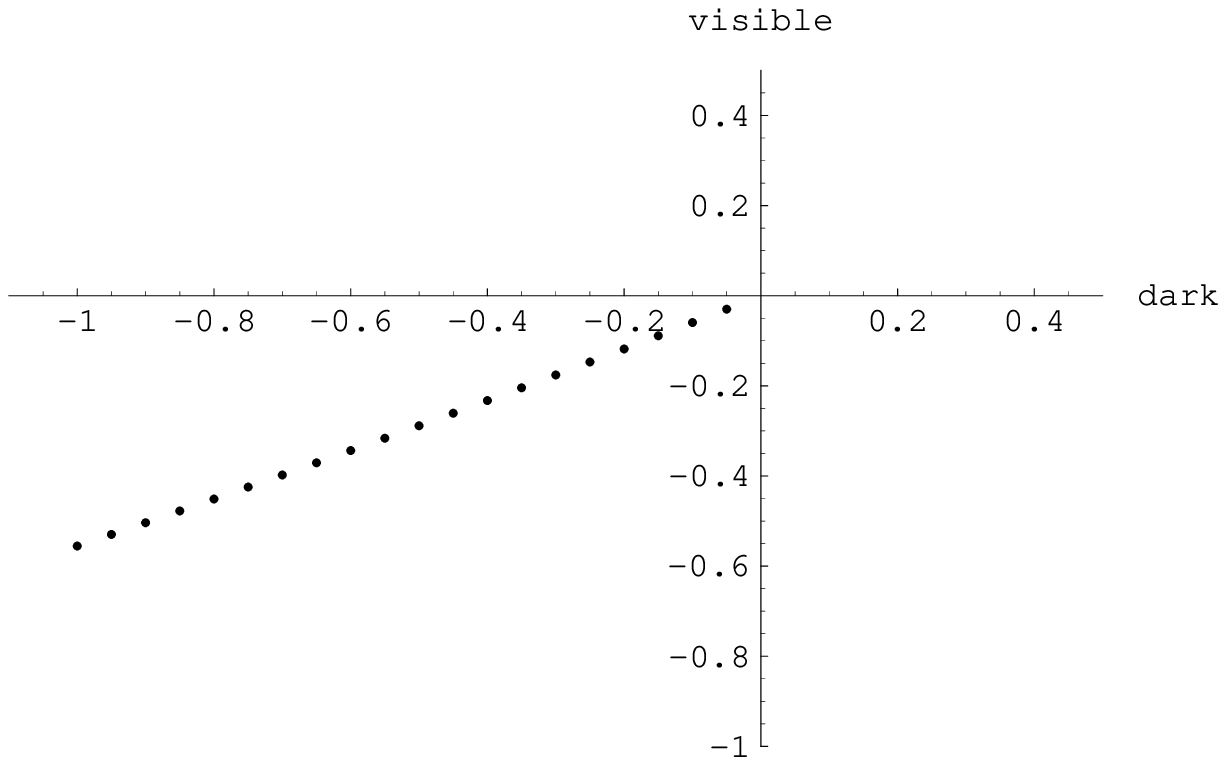,width=\hsize}
\end{minipage}
\caption{$\ln \rho_v$ vs $\ln \rho_d$ when the dark matter particle is 
light (500MeV)(see the left two plots) and heavy (4GeV)(see the right 
plot). For the left two plots, the visible component has the composition 
of 75\% $H_2$ + 25\% He (left) and 50\% $H_2$ + 50\% He (center), 
respectively. For the right plot, the visible component has
the composition 75\% $H_2$ + 25\% He.}
\label{ws}
\end{figure}
When this effect is taken into account, the correlation is still linear 
although the different relative contents obviously affect the slope of 
the line a little.
We conclude therefore that when the composite nature of the interstellar 
medium is taken into account the linear correlation between 
$\ln(\rho_v)$ and $\ln(\rho_d)$ is still preserved. 
The slope of this correlation is not very sensitive to the exact 
concentrations although the knowledge of the visible mass composition is 
needed to reliably extract the mass of dark matter particle from the 
slope of the correlation. From this and the above results we can estimate
the mass of dark matter particle, including the uncertainty in visible 
mass content, as $$\mu_d\approx(200 \sim 800)MeV = 500 \pm 300 MeV.$$ 

In more details, our error estimates include other uncertainties.
The statistical uncertainty in $\kappa$ we find to be
$\epsilon_1 \approx 20$\%. 
The uncertainty due to the effects of rotation can be estimated
by considering the case of spiral galaxies with and without correction
due to rotation in which case we find variance of about 
$\epsilon_2 \approx 20$\%. 
As we discussed in the beginning of this section, 
we assumed the local thermodynamic equilibrium between
dark and visible matter, {\it i.e.} $\beta_v(r)/\beta_d(r) \approx 1$. 
Although in the absence of any direct measurements of 
the properties of the dark matter it is simply impossible to estimate 
the uncertainty on $\beta_v(r)/\beta_d(r)$, 
we may hope that the variation due to this ratio in part revealed
itself through variation in measurements of $\kappa$ and thus may 
not be larger than the statistical uncertainty in $\kappa$.
Taking these uncertainties into account for given $\kappa$
we may relax the range to $\mu_v / \mu_d \approx 3.0 \sim 6.0$.
The uncertainty in the visible matter composition is more difficult to
estimate due to absence of reliable measurements as well as due to the fact
that this composition may vary greatly throughout the space. 
We estimate this uncertainty as due to the uncertainty in content of 
atomic and molecular $H$ in which case $\mu_v \approx (1200 \sim 2100) MeV$. 
Accumulating these uncertainties we get $\mu_d\approx (200 \sim 800) MeV$.

\section{Discussion and Conclusion}
We studied the mass distribution in the CL0024 galaxy cluster and RCs for 
the spiral galaxies from the thermodynamic point of view. 
We showed that the global visible and dark mass distribution in spiral 
galaxies is consistent with an almost isothermal Boltzmann distribution. 
We find that the $\mu/T$ factors of these distributions are close to 
a constant 
for a wide range of galaxy luminosities both for visible and 
dark matter. Consequently, we find that the ratio of $\mu/T$'s for 
visible and dark matter is a universal constant of approximately 
$\kappa \approx 4.4\pm0.8$. 
The same conclusion is also drawn from the analysis of mass distribution 
in the galaxy cluster CL0024.

The simplest interpretation of this result implies that the visible matter 
is approximately 4 times heavier than the dark matter.
If we narrow down the composition of visible matter as the 
one in the interstellar medium suggested in Ref.\cite{thronson},
{\it i.e.} 45\% H, 45\% H$_2$ and 10\% He, then we get that the dark matter 
is made of particles with mass of about 300MeV.
To our knowledge, there are no candidates of mass in this range within 
the current extensions of Standard Model. Massive neutrinos and axions are 
usually ruled out by the cosmological arguments of large structure
formation. Also, their experimental mass limits $<25$MeV for the 
heaviest $\tau$-neutrino \cite{hu} are well below our range.
The most favorable candidates for non-baryonic dark matter, such as 
Weakly Interacting Massive Particle (WIMP) with the mass anywhere between 
10GeV and 1TeV and the SUSY lightest particle, {\it e.g.} neutralino, with 
the mass above 30GeV \cite{khalil}, are also far out of the range yielded 
from our analysis. 
In fact, any existing dark matter particle candidate would result in 
either too stretched  (neutrinos and axions) or too shrunk (WIMP \& 
neutralino) halos and thus have difficulties explaining observed extent 
of dark halos. 
It is however interesting to note that the energy scale of 
300MeV is astonishingly close 
to the QCD scale $\Lambda_{QCD}$ and
constituent u-d quark masses. Whether our finding is indeed 
related to the QCD vacuum properties 
may deserve a further study. 

Because our analysis is based on the assumptions of the 
local statistical equilibrium of dark and visible matter and
the equal temperature between the visible and dark matter components,
we add some more comments here` risk of repeating ourselves.
The hypothesis of linear regime in Log-Log plots of dark vs. visible 
matter distributions is motivated by very elementary thermodynamics 
reasoning.
This regime, however, can be easily spoiled by various non-equilibrium and
rotation effects. Thus the validity of this hypothesis can be verified
only from the consistency of the final results. In our study we do find
consistent estimates for $\kappa$ from two tremendously different classes
of systems and we do find that the hypothesis of linear regime is
satisfactory in explaining the dark mass distribution in galaxy cluster
CL0024 as well as the URC in spiral galaxies of different luminosities.
Furthermore we find that the ratio of factors $\mu / T$ for
the visible and the dark components is
constant up to about 20\%. While it is still possible that the independant
variances in the temperature of visible and dark matter are all within our 
measurement uncertainty, this implies that $\beta_v$ and $\beta_d$ vary in 
a correlated way 
so that the ratio $\beta_v / \beta_d$ is preserved as a constant. 
Under these conditions
the most feasible explanation is that the ratio $\beta_v / \beta_d$
is unity due to thermodynamic equilibrium which implies the 
local statistical equilibrium and
the equal temperature between the visible and dark matter components.
Unfortunately we do lack crucial experimental evidences 
to provide the decisive argument here or to estimate properly uncertainty
caused by this assumption.
Yet, if the assumption of local thermal equilibrium is to be discarded, one
needs to suggest other physical mechanism that could provide constant observed
dark to visible temperature ratio other than 1. 

Also it may be still possible that $\mu$ in global 
Boltzmann distributions might be something other than the molar 
mass of the matter. Yet, the molar mass seems to be the most natural 
choice for $\mu$ and the consistency of the temperature of the visible matter distribution 
in this case shows that this choice is, indeed, the most viable variant. 
Finally, we also neglected the effect of rotation for the dark component 
(and 
in the case of the galaxy cluster CL0024 we also neglected the rotation 
effect of visible component) that could alter the value of $\kappa$. Such 
assumption, however, was natural for the given smooth elliptical form of 
dark halos. Still we estimate the uncertainty due to rotation from
considering spiral galaxies with and without effects of
rotation and include it in the total uncertainty of 
our final estimate for $\mu_d$.

To summarize, in our study, we found that the isothermal Boltzmann 
distribution is satisfactory in explaining observed visible and dark 
matter profiles in spiral galaxies.
The factors of these distributions are constant over a wide range of 
galaxy luminosities. The ratio of these factors for visible and dark 
matter is a constant. If this constant is related to the ratio of the 
molar mass of dark matter to that of the visible component, 
then it gives an estimate for the mass 
of dark matter particle: $\mu_d\approx (200 \sim 800)$ MeV.
The measurement in the galaxy cluster CL0024 along with the fact that the 
lightest component of the visible matter most significantly affects $\kappa$,
as was pointed out in Section IV, indicates that the mass of the dark 
matter particle is in fact closer to 300MeV, which appears astonishingly 
close to QCD energy scale.

\begin{appendix}
\section{The universal rotation curves of spiral galaxies}\label{appA}
The study in Ref.\cite{persic}, involving the samples of more than 1000 
spiral RCs, has shown that RCs grouped by luminosity follow certain 
universal profiles. These profiles were called Universal Rotation Curves 
(URC) and it was shown that they can be well fitted for a wide range of 
luminosities with a simple mass model including exponential disk and 
spherical halo.

According to the synthetic luminous mass profiles, the mass distribution 
in stellar disks can be well fitted 
by assuming $\rho_v\sim r^{-2} \exp (-3.2 r/r_{opt})$. 
The self-gravity on an infinitely thin disk with 
the surface mass density $I(x)\sim \exp (-3.2 x)$, ($x=r/r_{opt}$) 
yields an equilibrium circular velocity given by
\begin{equation}
v^2_v(x)=1.28 \beta v^2_{opt} x^2 (I_0(1.6x) K_0(1.6x) -I_1(1.6x) K_1(1.6x)),
\end{equation}
where $\beta = v^2_v(r_{opt})/v^2_{opt}=1.1 G M_v /v^2_{opt} r_{opt}$
and $I_0,I_1,K_0,K_1$ are the Bessel functions.
In \cite{persic}, this contribution is fitted with a simpler formula valid 
in the range $x\lesssim 2$: 
\begin{equation}
v^2_v(x)=v^2_{opt} \beta 1.97 \frac{x^{1.22}}{(x^2+0.78^2)^{1.43}}
\end{equation}
which we used in our calculations. Beyond $2r_{opt}$, the Keplerian 
regime is practically attained so that $v^2(r)_v=v^2_v(2r_{opt}) 
(2r_{opt}/r)$.

The contribution from a dark halo can be well represented by
\begin{equation}
v^2_d(x)=v^2_{opt} (1-\beta) (1+\alpha^2) \frac{x^2}{x^2+\alpha^2}
\end{equation}
with $\alpha$ being the `velocity core radius' normalized to $r_{opt}$. 
The URC profile then can be very well fitted by assuming
\begin{equation}\label{fits}
\begin{array}{c}
v_{URC}^2(r)=v^2_d(r)+v^2_v(r), \\
\beta=0.72 + 0.44 \log (L/L_*), \\
\alpha=1.5 (L/L_*)^{1/5},
\end{array}
\end{equation}
with $L$ being the I-band luminosity $L/L_*=10^{-(M+21.9)/5}$. 
Eq.(\ref{fits}) reproduces the URC for various luminosities M in the range 
from -18.5 to -23.2 within their rms. Small $r$ behavior of the URC is 
dictated by stellar disk and is close to $v^2\sim r$, while at large 
distances URC is dominated by dark halo contribution (especially for 
smaller galaxies) and is close to constant.

\section{Interesting scaling properties of CL0024 mass profile}\label{appB}
As was pointed out the observed mass density for dark matter is 
typically inconsistent with the simulations with the cold dark matter (CDM) 
model. Another naive try would be to consider an isothermal gas cloud model.
If one neglects the rotation, fairly simple result for the mass density 
profile can be obtained. This case is known as the Lane-Emden equation 
\cite{natarjan,emden} 
\begin{equation}\label{laneemden}
\frac{d^2 \omega}{dy^2}+\frac{2}{y}\frac{d\omega}{dy}=e^{-\omega},
\end{equation}
where y is the distance normalized to $r_0=\sqrt{\sigma^2/4\pi G\rho_0}$ 
($\rho_0$ is the mass density at the origin, {\it i.e.} $r=0$)
and $\sigma$
is the constant velocity dispersion (related to the temperature of the 
sphere $T$ and the particle mass $\mu$ by $\sigma^2=kT/\mu$), 
so that the mass density is given by
\begin{equation}
\rho(r)=\rho_0 e^{-\omega (r/r_0)}.
\end{equation}
The isothermal sphere solution has a soft core which is 
qualitatively consistent with
the observed mass distribution in CL0024 \cite{shapiro}, 
however, as more detailed study shows, 
it falls off too rapidly and as such is not quantitatively 
acceptable.

If one agrees, however, that isothermal sphere is good in describing
some global features of the CL0024 profile, one might expect then 
to see some of major features of the isothermal solution in experimental 
profile.
One of such features is the specific form of the solution, {\it i.e.}
$\ln ( \rho_i(r)/\rho_i(0) ) \sim \omega_i(r/r_{0,i})$. On the Log-Log 
plot this would mean that $\ln ( \rho_i/\rho_i(0) ) \sim f(z-z_i)$ where 
$z=\log (r)$.
Guided by this idea we compare the experimental mass profiles for 
the visible and dark components as well as for the total mass. One can see 
that, indeed, 
the mass profiles are aligned pretty well with simple horizontal shift. 
We find that the visible component profile is shifted on the Log-Log plot
with respect to the dark component profile by $\Delta z\approx-0.62$, 
while the total mass profile is shifted by $\Delta z\approx0.22$. The 
visible component mass profile is shifted with respect to the total mass 
profile by $\Delta z\approx-0.4$.
\begin{figure}[htb]
\centering
\begin{minipage}[c]{0.3\hsize}
\epsfig{file=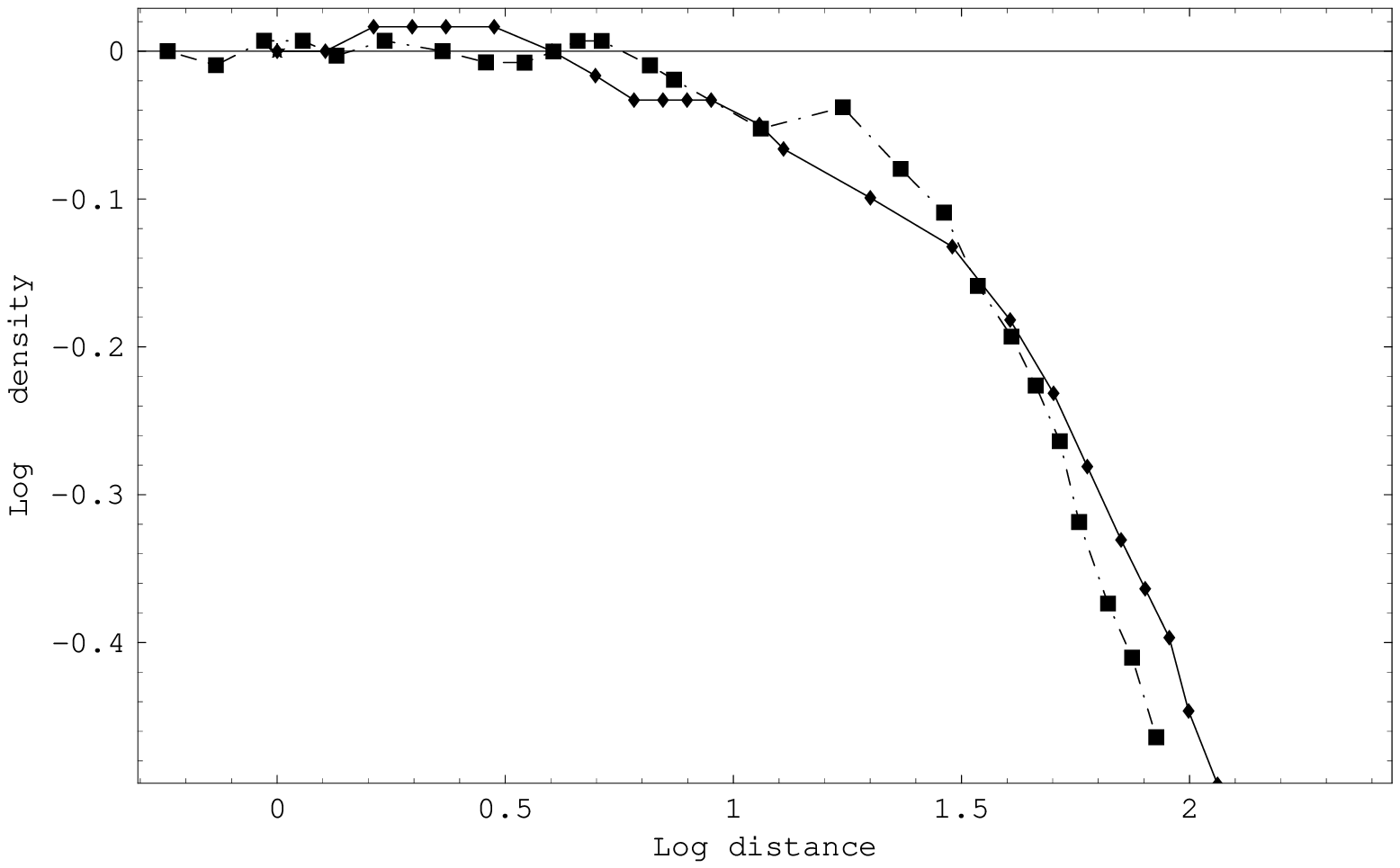,width=\hsize}
\end{minipage}
\hspace*{0.5cm}
\begin{minipage}[c]{0.3\hsize}
\epsfig{file=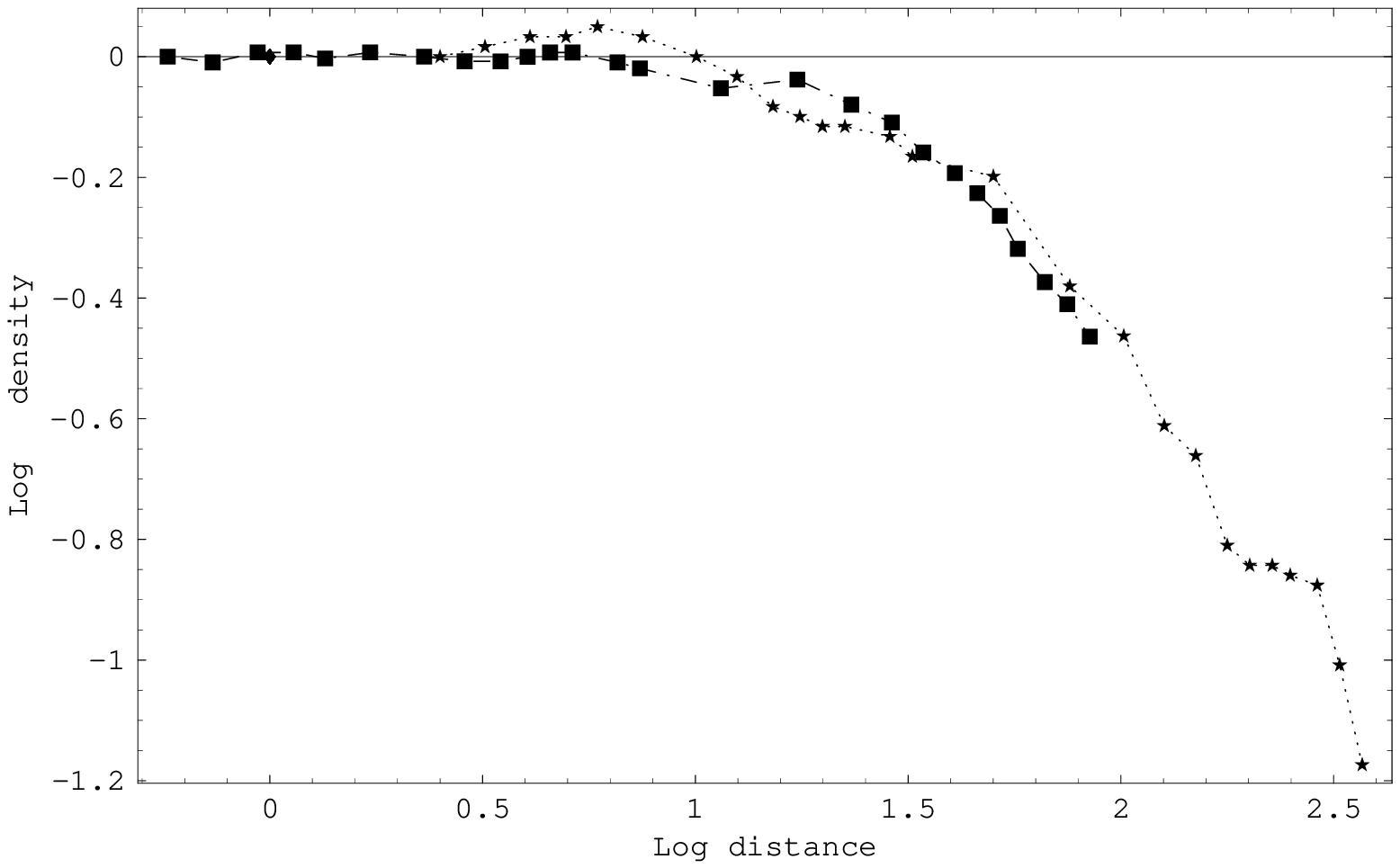,width=\hsize}
\end{minipage}
\hspace*{0.5cm}
\begin{minipage}[c]{0.3\hsize}
\epsfig{file=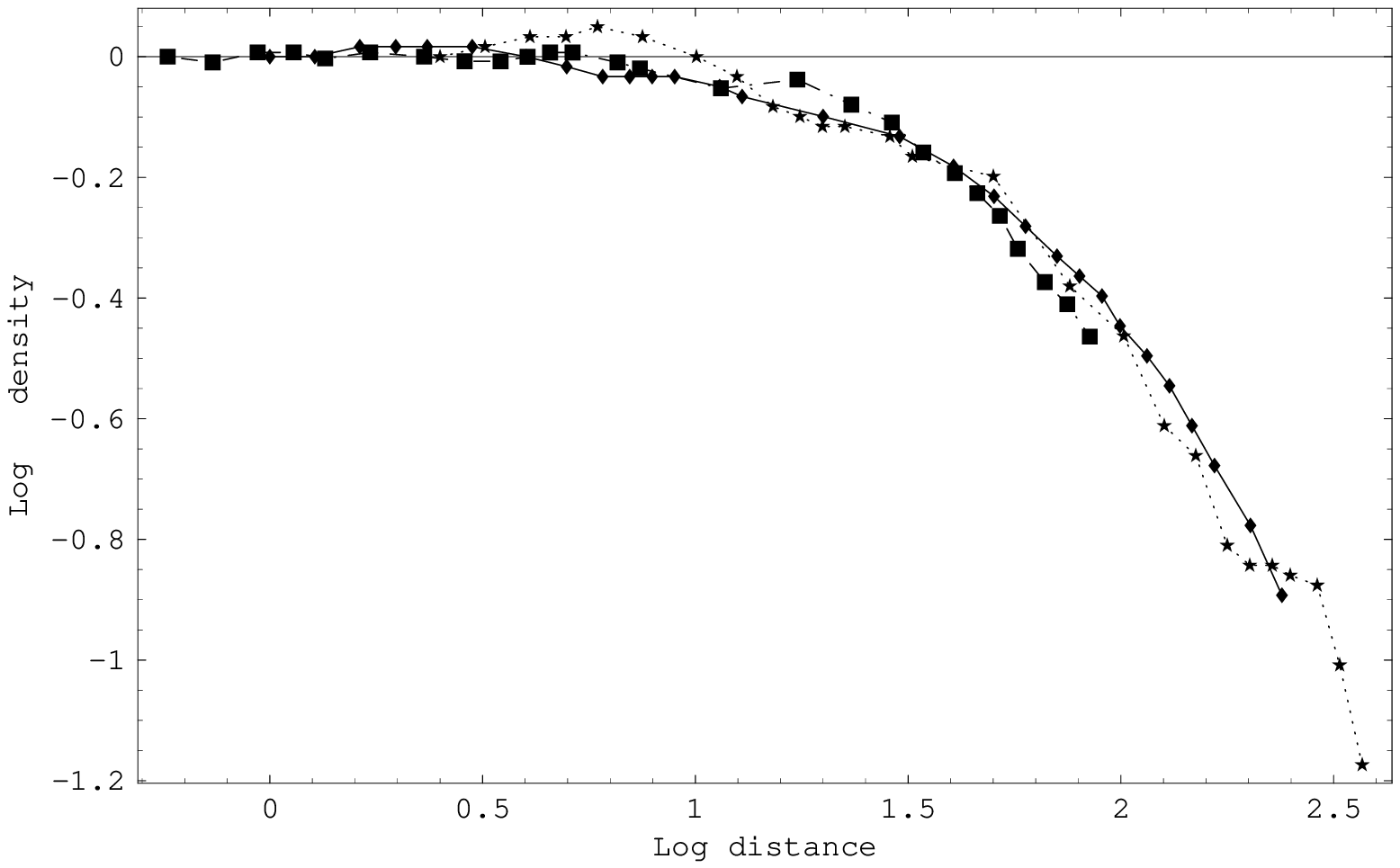,width=\hsize}
\end{minipage}
\caption{Alignment by horizontal shift of the total mass (left) and
visible mass profiles (center) with the dark mass profile. Alignment
of all three mass profiles is presented in the right panel.}
\label{align1}
\end{figure}
We believe this observed similarity can be understood in terms of the 
dimensionless variables of possible configurations of the rotating (and 
possibly not isothermal) self-gravitating gas.

\end{appendix}


\begin{thebibliography}{99}
\bibitem{tyson} J.A. Tyson {\it et al.}, astro-ph/9801193. 
\bibitem{persic} M.Persic, P.Salucci, F.Stel:MNRAS, 281, 27 (1996).
\bibitem{straumann} N. Straumann, astro-ph/0108255. 
\bibitem{wu} X-P. Wu, T. Chiueh, L-Z. Fang and Y-J. Xue, asto-ph/9808179. 
\bibitem{kneib} J-P. Kneib, astro-ph/0009385. 
\bibitem{abdelsalam} H. M. AbdelSalam, P.Saha, astro-ph/9806244. 
\bibitem{bezecourt} J.Bezecourt \& all, astro-ph/0001513.
\bibitem{white} M. White, L. Hernquist, V. Springel, astro-ph/0107023.
\bibitem{shapiro} P.R. Shapiro I.T.Iliev astro-ph/0006353.
\bibitem{navaro} J.Navaro, C.Frenk, S.White; ApJ, 462, 563 (1996).
\bibitem{hatsopoulos} G.N.Hatsopoulos, J.H.Kennan, 
{\it Principles of general thermodynamics},
Wiley, pp.513-515, 1965. 
\bibitem{votyakov} E.V. Votyakov \& all, Phys. Rev. Lett. {\bf 89}, 031101 (2002); cond-mat/0202140.
\bibitem{vega1} H.J. de Vega, N.Sanchez, Nucl. Phys. {\bf B625},460-494 (2002); astro-ph/0101567.
\bibitem{combes} F.Combes, astro-ph/0206126.
\bibitem{albada} T.S. van Aldaba, J.S.Bahcall, 
K.Begeman, R.Sancisi:ApJ, 295, 305 (1985).
\bibitem{salucci} P.Salucci, A.Borriello, astro-ph/0203457.
\bibitem{thronson} {\it The Interstellar Medium in Galaxies},ed. H.A.Thronson
and J.M.Shuller, 1990, Proceedings of 2nd Tetons Conference, Kluwer
Publisher, 556 pages.
\bibitem{vega} H.J. de Vega, J.A. Siebert, Phys. Rev., {/bf E66},016112 (2002); astro-ph/0111551.
\bibitem{hu} W.Hu, D.Eisenstein, M.Tegmark, Phys. Rev. Lett. {\bf 80}, 5255-5258 (1998); astro-ph/9712057.
\bibitem{khalil} S.Khalil, C.Munoz, Contemp. Phys. {\bf 43}, 51-62 (2002); astro-ph/0110122.
\bibitem{natarjan} P. Natarajan, D. Lynden-Bell astro-ph/9604084.
\bibitem{emden} R. Emden, 1907 in Gaskugeln-Anwendungen der Mechan. 
Warm theorie. Druck und verlag Von B.G. Teubner, Leipzig, p.129.
\end{thebibliography}
\end{document}